\documentstyle[11pt,epsfig]{article}
\newcommand{\ba}{\begin{array}}
\newcommand{\ea}{\end{array}}
\newcommand{\bd}{\begin{displaymath}}
\newcommand{\ed}{\end{displaymath}}
\newcommand{\be}{\begin{equation}}
\newcommand{\ee}{\end{equation}}
\newcommand{\bea}{\begin{eqnarray}}
\newcommand{\eea}{\end{eqnarray}}

\def\q2 {q^2}

\begin{document}
%THE TEXT STARTS HERE

%\begin{flushright}
%{\large MRI-PHY/P990928 \\
%\large SISSA 115/99/EP \\ 
%September, 1999  
%\\ hep-ph/9910296}
%\end{flushright}
%\vskip1cm

\begin{center}
{\Large\bf Tevatron signatures\\[1.1ex] 
of an R-parity violating supersymmetric theory}\\[7mm]
Aseshkrishna Datta %\footnote{E-mail: asesh@mri.ernet.in}  
and 
Biswarup Mukhopadhyaya %\footnote{E-mail: biswarup@mri.ernet.in}
\\ 
{\em Mehta 
Research Institute,\\
Chhatnag Road, Jhusi, Allahabad - 211 019, India}, 
\\[4mm]
Francesco Vissani %\footnote{E-mail: vissani@lngs.infn.it}
\\
{\em INFN, Laboratori Nazionali del Gran Sasso, I-67010 Assergi (AQ), Italy;\\
Scuola  Internazionale Superiore di Studi Avanzati (SISSA), Trieste, Italy} 
\end{center}

\begin{abstract}
We show that an R-parity violating supersymmetric scenario which can 
account for the atmospheric $\nu_\mu$ deficit has testable signals
at the Fermilab Tevatron with upgraded energy and luminosity. The 
explanation of neutrino masses and maximal $\nu_\mu$--$\nu_\tau$ 
oscillation in terms of bilinear
R-violating terms in the superpotential associates comparable numbers
of muons and tau's resulting from decays of the lightest neutralino.
We show that this should lead to like-sign dimuons and ditaus
with substantial rates, in a form that separates them from standard model 
backgrounds and other signals of supersymmetry. One here also has the 
possibility of completely reconstructing the lightest neutralino. 
\end{abstract}
\vskip 5pt

\setcounter{footnote}{0}

\def\baselinestretch{1.3}

Existing data \cite{neutexp} strongly point 
towards nondegenerate, massive neutrinos and
large mixing between $\nu_\mu$ and $\nu_\tau$ 
(or, perhaps, a sterile state to which $\nu_\mu$ oscillates).
To decide on which extension of the standard model
fits this picture best, one also needs to think of
other testable consequences of any suggested scenario.
In this spirit, here we predict some signals of one
such candidate theory \cite{mrv}:  
a supersymmetric (SUSY) scenario \cite{habkane}
where R-parity breaking \cite{rpv} bilinear terms 
\cite{BILI,chun} mix neutrinos and 
neutralinos \cite{SVEV}, and thence lead to
massive neutrinos.

Admitting of lepton number violation, the superpotential for the
minimal SUSY standard model (MSSM) can be extended to include the 
following terms:
\begin{equation}
W_{\not L} =  \epsilon_i {\hat L}_i {\hat H}_2 +            
\lambda_{ijk} {\hat L}_i {\hat L}_j {\hat E}_k^c +
\lambda_{ijk}' {\hat L}_i {\hat Q}_j {\hat D}_k^c 
\end{equation}  
Here we take the most econonomic aproach and 
consider only the bilinear terms in the 
second and third generations, {\it viz.} $\epsilon_{2,3} L_{2,3} H_2.$ 
These terms can be responsible for generating one tree-level neutrino mass;
also, the retention of the bilinears in the 
second and third generations ensures
$\nu_\mu - \nu_\tau$ mixing of the kind suggested by the atmospheric
neutrino data \cite{mrv}.

Performing a rotation  
\cite{chun}; \cite{datta} we remove them from the superpotential, 
and reabsorb them in $\mu' H_1' H_2$
(where $\mu'= (\mu^2 + \epsilon^2_2 + \epsilon^2_3)^{1/2},$ and $H_1^{'}$ is a 
linear combination of $H_1$, $L_2$ and $L_3$). 
But the effects of  $\epsilon_2$ and $\epsilon_3$ are still manifest; 
apart from newly produced trilinear terms, 
they show up through the scalar potential,
where non-zero sneutrino \emph{vev}'s 
$v_\mu$ and $v_\tau$ are in general induced. 
These \emph{vev}'s give rise to neutrino-gaugino cross terms,  
that change the structure of the neutralino mass matrix
\begin{equation}
\frac{1}{2}
\Psi^t C^{-1}
 \left( \begin{array}{cccccc}
  0 & -\mu' & \frac {gv_2} {\sqrt{2}} &
  -\frac {g'v_2} {\sqrt{2}} & 0 & 0 \\
  -\mu' & 0 & -\frac {gv'_1} {\sqrt{2}}
       & \frac {g'v'_1} {\sqrt{2}} & 0 & 0 \\
 \frac {gv_2} {\sqrt{2}} & -\frac {gv'_1} {\sqrt{2}} & M & 0 & -\frac
 {gv_\tau}
 {\sqrt{2}} & -\frac {gv_\mu} {\sqrt{2}} \\
 -\frac {g'v_2} {\sqrt{2}} & \frac {g'v'_1} {\sqrt{2}} & 0 & M' &
  \frac {g'v_\tau} {\sqrt {2}} & \frac {g'v_\mu} {\sqrt {2}} \\
  0 & 0 & -\frac {gv_\tau} {\sqrt {2}} & \frac {g'v_\tau} {\sqrt {2}} &
  0 & 0  \\
  0 & 0 & -\frac {gv_\mu} {\sqrt {2}} & \frac {g'v_\mu} {\sqrt {2}} &
  0 & 0
  \end{array}  \right) \Psi, 
\ \ \ \ \ \mbox{with }
\Psi=\left(
\begin{array}{c}
\tilde{H_2}\\ 
\tilde{H_1'}\\ 
-i\tilde{W}_3\\ 
-i\tilde{B} \\
\nu_\tau\\
\nu_\mu 
\end{array}
\right)
\end{equation}
$M$ and $M'$ are the ${\rm SU(2)}$ and ${\rm U(1)}$ gaugino 
masses, and $v'_1 = \langle H'_1 \rangle$, $v_2 = \langle H_2 \rangle .$
Thence, the state
\begin{equation}
\nu_3 = \nu_\tau \cos \theta + \nu_\mu \sin \theta  
\ \ \ \ \mbox{ where }\tan\theta=v_\mu/v_\tau
\end{equation}
acquires a tree-level mass. 
For  ${m_Z}\ll {\mu'},$  this mass is
\begin{equation}
m_{\nu_3}\approx 
-\frac{\bar{g}^2 (v_\mu^2+v_\tau^2)}{2\ \bar{M}}\times
\frac{\bar{M}^2}{M M' - m_Z^2\ \bar{M}/\mu' \ \sin 2\beta }
\end{equation}
where  
$\bar{g}^2 \bar{M}=g^2 M'+{g'}^2 M$, and $\tan \beta=v_2/v_1'.$ 
The parameter $v_0=(v_{\mu}^2 + v_{\tau}^2)^{1/2}$ is
a basis-independent measure of R-parity violation; if
$v_0\sim 100$ keV 
(say, for $\mu'=-500$ GeV, 
$\tan \beta=~5$ and $m_{\tilde{\chi}_1^0} 
\simeq 100$ GeV) the atmospheric neutrino problem 
can be explained in terms of 
oscillations\footnote{Loop effects split the residual degeneration 
in the neutrino mass matrix; this permits to address 
the deficit of solar $\nu_e$ \cite{bahcall} in the context of the 
present model (see also \cite{bakira}).}. 
The decay of the lightest neutralino $\chi_1^0$ 
(assumed to be the LSP--lightest SUSY particle) 
could provide the crucial test of this model:
In fact, if enough massive, $\chi_1^0$
has two-body decays 
of the type
$\chi_1^0 \longrightarrow l W$ and $\chi_1^0 \longrightarrow \nu_{l} Z ;$
$l=\mu , \tau$ (Roy-Mukhopadhyaya in \cite{BILI}; \cite{mrv,datta,BILEX}).

The production and cascade decays of superparticles at a hadronic collider
(till the LSP is reached) is expected to be controlled by R-parity conserving
interactions.  Thus all SUSY processes at the
Tevatron will end up in $\chi_1^0$ pairs.
Two-body decays of a neutralino will 
lead often to lepton-$W$ final states;
a conservative estimate predicts 
B($\chi_1^0 \to \mu W)\approx$
B($\chi_1^0 \to \tau W)\approx 35$\% 
for maximal mixing angle \cite{mrv}. 
Half of these events contain like-sign dileptons (LSD) 
due to the Majorana  character of neutralinos \cite{dpmona}. 
It is this $real~ W + LSD$ signal (with 
correlated numbers of dimuons and ditaus, but 
a depletion of dielectrons) 
that can help us testing the proposed model.
Such a signal is  of particular
interest for the CDF detector, sensitive to LSD's;
it has already been used to look for R-parity violation of
other types with $107~pb^{-1}$ data sample obtained
at $\sqrt{s} = 1.8$ TeV \cite{cdf}.
Here we present our results for the upgraded Tevatron running 
at $\sqrt{s}=2$ TeV, with a luminosity of 10 fb$^{-1}$.

We assumed that the only  QCD process that leads to
LSP pairs is the pair-production of squarks
(of five degenerate flavours)
whose lower mass limit is approaching 300 GeV \cite{d0lim}. 
The effective decoupling of the
gluino can be indeed derived from the 
condition that the lightest neutralino 
heavier than the $W,$ assuming gaugino mass unification
(that we use to reduce the number of parameters).
We retained all possible cascade channels
in our calculation, and took particular note of the small but 
non-negligible contributions from $\chi_2^0 \chi_1^{\pm}$ 
and $\chi_1^{+} \chi_1^{-}$ 
($\chi_2^0$ and $\chi_1^{\pm}$ are the second 
lightest neutralino and the lighter chargino). 
Also, we have {\it not} restricted ourselves to a 
supergravity (SUGRA) scenario \cite{tata} but 
considered squark and slepton masses 
as free parameters within existing experimental bounds
(this is a common practice in Tevatron SUSY analyses;
see also Ref.\ \cite{mssm0}).

The signals involving leptons get enhanced 
considerably if one remembers that the cascades leading to LSP pairs 
also produce single-or multi-leptons simultaneously. For example,
heavier neutralinos
(mainly $\chi_2^0$) and the lighter chargino ($\chi_1^{+}$) can cause 
cascades where sleptons giving large
contributions if they are so light as to be produced on-shell.
On the whole, LSP pairs are
in general produced in final states of the following types:
\begin{equation}
\chi_1^0~ \chi_1^0~+~0,1,2 \mbox{ or } 3\ l~+~X  
\ \ \ \ \mbox{ with $l=e,\mu,\tau$}
\end{equation}
%\begin{itemize}
%\item $\chi_1^0~ \chi_1^0~+~X$
%\item $\chi_1^0~ \chi_1^0~+~l~+~X$
%\item $\chi_1^0~ \chi_1^0~+~ll~+~X$
%\item $\chi_1^0~ \chi_1^0~+~lll~+~X$
%\end{itemize}
Charged leptons can be produced 
in the cascade, or in the decay $\chi_1^0$-pair;
$X$ denotes other states, possibly 
produced in association (jets, neutrinos).

%\vskip 15pt
\begin{table}[bht]
\begin{center}
\begin{tabular}{|cccc|c|c||c|c|c|c||c|c|c|}
\hline
$\tan \beta$ & $M_2$ & $m_{\tilde{g}}$ & $m_{\chi_1^0}$ & $m_{\tilde{q}}$ &
$m_{\tilde{l}}$ & $\sigma_0$ & $\sigma_1$ & $\sigma_2$ & $\sigma_3$ &
$\sigma_{1l}$ & $\sigma_{2l}$ & $\sigma_{3l}$ \\
\hline
 & & & & & 150 & .218 & .233 & .078 & .012 & .045 & .013 & .021 \\
 & & & & 300 & 225 & .479 & .118 & .017 & .001 & .002 & .001 & .001 \\
% & & & &  & {\bf 225} & {\bf .319} & {\bf .082} & {\bf .012} & {\bf .001}
% & {\bf .002} & {\bf .001} & {\bf .001} \\
3 & 200 & 698 & 102 &  & 300 & .477 & .117 & .019 & .001 & .003 & .001 & .002 \\
\cline{5-13}
 & & & & & 150 & .016 & .023 & .006 & .001 & .045 & .018 & .021 \\
 & & & & 400 & 225 & .086 & .034 & .004 & .000 & .003 & .002 & .001 \\
% & & & &  & {\bf 225} & {\bf .067} & {\bf .029} & {\bf .003} & {\bf .000}
% & {\bf .002} & {\bf .002} & {\bf .001} \\
 & & & &  & 300 & .085 & .033 & .005 & .000 & .003 & .002 & .002 \\
\hline
\hline
 & & & & & 150 & .163 & .217 & .119 & .029 & .031 & .023 & .075 \\
 & & & & 300 & 225 & .496 & .124 & .036 & .002 & .001 & .002 & .004 \\
% & & & &  & {\bf 225} & {\bf .337} & {\bf .089} & {\bf .025} & {\bf .001}
% & {\bf .001} & {\bf .002} & {\bf .004} \\
30 & 200 & 698 & 97 &  & 300 & .518 & .126 & .015 & .001 & .003 & .002 & .002 \\
\cline{5-13}
 & & & & & 150 & .012 & .021 & .010 & .002 & .031 & .031 & .075 \\
 & & & & 400 & 225 & .111 & .048 & .014 & .000 & .001 & .003 & .005 \\
% & & & &  & {\bf 225} & {\bf .090} & {\bf .041} & {\bf .012} & {\bf .000}
% & {\bf .001} & {\bf .003} & {\bf .004} \\
 & & & &  & 300 & .120 & .048 & .005 & .000 & .003 & .003 & .002 \\
\hline
\end{tabular}
\end{center}
{\small Table 1: Lowest order (LO) cross sections in pb at
$\sqrt{s}=2$ TeV for various final states and model parameters  
($\mu'=-250$ GeV). 
$\sigma_n$'s ($n=~0,1,2,3$) are the cross sections for
$n$-lepton + jets + $\chi_1^0 \, \chi_1^0$ + $E\!\!\! /$ 
(carried by neutrinos) final states; $\sigma_{nl}$ is the cross
section for hadronically quiet $n$-lepton event. 
The numbers shown as .000 are insignificant up to three 
decimal places. We used CTEQ-4L 
parametrization (from PDFLIB, \cite{pdflib}) evaluated
at $Q=(m_i+m_j)/2$  ($m_{i,j}$ are the masses of the 
particles produced in the hard scattering).
Next-to-LO corrections enhance
  $\sigma_0$ and $\sigma_1$
  by $\sim 20$\% .}
\end{table}

\vskip 8pt

The sample results\footnote{Our results are cross-checked in the appropriate 
limits against, for example, the calculation in \cite{spira-berger}.} 
shown in Table 1 demonstrate that, of the 
different final states mentioned above,  the `pure jet + LSP pair' and  the 
`single lepton' final state have the highest rates in general, the dominant
channel being determined by whether a slepton is light enough to be produced
on-shell.
%\footnote{These numbers are obtained at the leading order 
%on setting the QCD 
%renormalisation scale at the average mass of the final state particles. 
%They are subject to enhancement at next-to-leading order \underline{(NLO)}
%by a $K$-factor of 
%about 1.2. 
%On the other hand, setting the renormalisation scale to the
%subprocess centre-of-mass energy reduces $\sigma_0$ and $\sigma_1$ by
%15 - 33\%, while the other cross-sections are relatively unaffected. 
%Also,
%our results are cross-checked in the appropriate limits against, 
%for example, the calculation of squark pair production in reference 
%\cite{spira-berger}.} 
These rates are  
followed by that of the di-lepton final state. In the latter 
cases, however, our signals can receive significant
contributions because, if one of the leptons ($\mu$ or $\tau$) 
pairs up with one of identical flavour and
sign coming from LSP decay, the overall rate gets suppressed only by
a single power of the LSP branching fraction rather than its square. Thus,
by reconstructing {\em one} $\chi_1^0$ and letting the other one decay in any
allowed channel, one may get an enhancement by about 
a factor of 6 so long as it is enough to
look for just one reconstructed $W$ in the final state. The most convenient 
mode for reconstructing the $W$ and hence, checking the mass-shell condition 
of $\chi_1^0$ seems to be $\chi_1^0 \longrightarrow lW$ ($l=\mu,\tau$) 
followed  by $W \longrightarrow jj$, since all the decay products are
visible here.

The standard model backgrounds to this kind of a 
final state ($W+LSD$) are quite suppressed. This is because the 
neutralino decay length in our case can be as large as between 0.1 $mm$ and 
1 $cm$ \cite{mrv,chun} when the decaying neutralino is of mass around 100 
GeV or above. Such decay gaps result from the smallness of the
R-parity violating  coupling driving decays of the LSP. This coupling 
is determined by the basis-independent parameter $v_0$
which also controls the tree-level neutrino mass. The resultant 
displaced vertex-- to which the $W$ needs to be reconstructed--  will
distinguish our signals. For example, LSD backgrounds 
from $t\bar{t}$-production followed by semileptonic decay of one of the bottom
quarks produced in top-decay can be 
eliminated by proper identification of the primary and secondary vertices 
together with suitable isolation cuts on the leptons.
Though most MSSM cascades 
potentially faking our signals are
suppressed at some stage or other, there are processes like
$p\bar{p} \longrightarrow \tilde{b} \tilde{b}^{*}$ followed by
decays of the $b$-squarks to a top, where both $W$'s and like-sign dileptons 
can be observed. However, the decay gap typical of our
scenario can cause the latter to stand out. 
When R-parity is violated by trilinear terms, LSP's produced via 
the decay $\chi_1^+ \longrightarrow \chi_1^0 W$, followed by three-body 
leptonic decay of each LSP, can in principle give rise to $LSD+W$,
where a similar decay gap as above can occur if the trilinear R-violating
couplings are appropriately small. In order to avoid faking by signals of 
this kind, one could test whether one lepton in the $LSD$ pair and the 
$W$ originate from the same displaced vertex. In addition, here as well
as for the  $t\bar{t}$ backgrounds mentioned above, one can use the 
neutralino mass scale condition to suppress the backgrounds.

Thus the finally suggested signals are of the form
\begin{equation}
\mbox{\em like-sign dimuons/ditaus 
+ displaced vertex 
+ a real W paired with one $\mu/\tau.$}
\end{equation}\nonumber 
The condition of 
having one displaced vertex 
forces us to leave out 
those events where both leptons have their
origins in cascades rather than in 
$\chi_1^0$ decays.

In figures 1-3 we show some plots of the predicted number of signal events  
(of both like-sign dimuons and ditaus) expected at the Tevatron Run II 
($\sqrt{s}=2$ TeV, with an integrated luminosity of 10 fb$^{-1}$),
once the above set of criteria are specified. 
The Higgsino 
mass parameter $\mu$ is taken to be $-250$ GeV in all the cases. 
No drastic effect
in the nature of these curves are observed due to variations 
in $\mu$.  
A common low-energy slepton mass has been
assumed in each case. In addition, the physical stau masses are determined 
by left-right mixing depending on the value of tan$\beta$.
Each curve is truncated at
the point where the lightest neutralino ceases to be lighter than 
{\em all} the sleptons. 

Let us discuss the most prominent features of these results:\newline
(i) For slepton masses of 150 GeV, leptonic final states 
can be observed via decays of on-shell sleptons produced in the preceding
decays of neutralinos or charginos in the cascades. 
For low $\tan \beta$, like-sign dimuon and ditau events rates are 
closely  comparable. As $\tan \beta$ increases,
mixing in the stau sector lowers the mass of one physical state 
while still keeping it consistent with the current search limits\cite{L3stau}.
This results in larger event rates with ditaus and a consequent 
splitting between the dimuon and ditau curves as seen from figures 1, 2 and 3. 
Also, $\mu\tau$-events with similar kinematic characteristics as our
already described signals are expected in this scenario. The $\mu\tau$-type
LSD event plots should correspond approximately to the sum of the dimuon 
and ditau curves for each combination of parameters.  \newline
(ii) With the slepton masses on the higher side (225 GeV and 300 GeV), the 
event rates are mostly controlled by the all-jet channel for relatively
low masses of the lightest neutralino. In these regions, for low $\tan \beta$,
the numbers of muonic and tau events are very close together, and 
are relatively insensitive to the above variation in the 
slepton masses. However, with large enough gaugino masses, two-body decays of 
$\chi_2^0$  and $\chi_1^+$ in the cascades become possible, 
whereby  rates for the single-and dilepton +
$\chi_1^0 - pair$ events get enhanced.  This causes both the muonic and 
tau-events to go up comparably for lower values of $\tan \beta$ for which
there is no substantial mass-splitting between the smuon and stau mass
eigenstates. 
For larger values of $\tan \beta$ ($\tan \beta=10,30$) in this region
(see figures 2 and 3) there is a rather early onset of the two-body decays 
of $\chi_2^0$ and $\chi_1^+$ leading to a sharp enhancement in tauonic events
while a relatively minor rise of this type is noticed for the corresponding
muonic events at a larger value of gaugino masses. The difference is attributed
primarily to the fact that for same $\tan \beta$ the lighter $\tau$ mass
eigenstate is less massive than its muonic counterpart. This in turn implies
that the two-body decays of gauginos involving 
muons open up for somewhat higher
gaugino masses which suppress the signals from the production level as well. 
The rise in the events rates due to the above effects is offset by the
fall in the rate of $\chi_2^0$ and  $\chi_1^+$ production when the masses  
of the latter (and therefore that of $\chi_1^0$) go on increasing. That is
why the curves showing the tau-rates as shooting up also show a
fall following a peaking behaviour. \newline
(iii) However, for a slepton mass of 300 GeV these features are not clear 
from 
the graphs. The relevant cross sections surely get diminished due to heavier
sleptons and LSPs in the processes leading to a smaller number of events. 
These, when plotted with the curves for lower slepton masses, 
fail to attain the
required resolution due to obvious reasons.

Our results indicate that rather copious numbers of uncut 
signal events\footnote{Actual analyses need to consider 
ID-efficiencies  for $\mu,$ $\tau$ and $W$'s, and isolation 
and $p_T$ cuts for $\mu,$ $\tau$.},
unlikely to be faked by backgrounds, are expected over 
a large part of the parameter space. 
Comparing the upper and lower panels
of each of of figures 1,2 and 3 it is seen that the number of events fall 
with a rise in the squark masses, since, as we have already mentioned, 
squark pairs form the main source of cascades here. Apart from that the 
general features of the events remain unaltered for fixed $\mu$ and 
$\tan \beta$.

In fig.\ 4 we show the muon $p_T$-distributions in some sample 
cases. We have separately plotted the distributions coming from
decays of the LSP ({\it i.e.}\ the ones emerging from displaced
vertices) and those produced in earlier stages of the cascades ({\it i.e.}\
the ones that would be there even if R-parity were conserved). 
Increasing the LSP mass, the leptons from the cascades are
less important for the overall signal strength, while the 
muons originating from the LSP tend to be harder. 
We have used a slepton  mass of 150 GeV in these 
plots; the cascades, as demonstrated in Table 1, are considerably less
significant for higher slepton masses. These  conservative 
estimates show that a large number of muons
in each case survive an $p_T$-cut of the order of 15 GeV, 
usually adopted in Tevatron experiments \cite{et12}. 
Taking the overall detection efficiencies 
into account, we expect that well above 50~\% 
of the events predicted  in figures 1-3 
can be salvaged after the experimental 
cuts employed at the Tevatron.

In conclusion, the decays of the lightest 
neutralino can have quite distinctive signatures in the
considered R-parity violating scenario. 
The characteristic signals are final states 
comprising of like-sign dimuons and ditaus, 
together with a real $W,$ with a measurable 
decay gaps (${\cal O}(1 mm)$ or larger).
These criteria eliminate any backgrounds, 
and distiguish the signal from those of other 
R-parity breaking models. 
An effective $W$-reconstruction 
leads to observation of the lightest
neutralino mass peak. These considerations
motivate to a certain extent the urge 
for further refinement of the techniques 
of such reconstruction as well as of the resolution of 
displaced vertices at Run-II of the Tevatron.

We thank V. Ravindran for useful discussions.

\newpage

\begin{figure}[htb]
\begin{center}
\vskip-7.5cm
\hskip-1cm\centerline{\epsfig{file=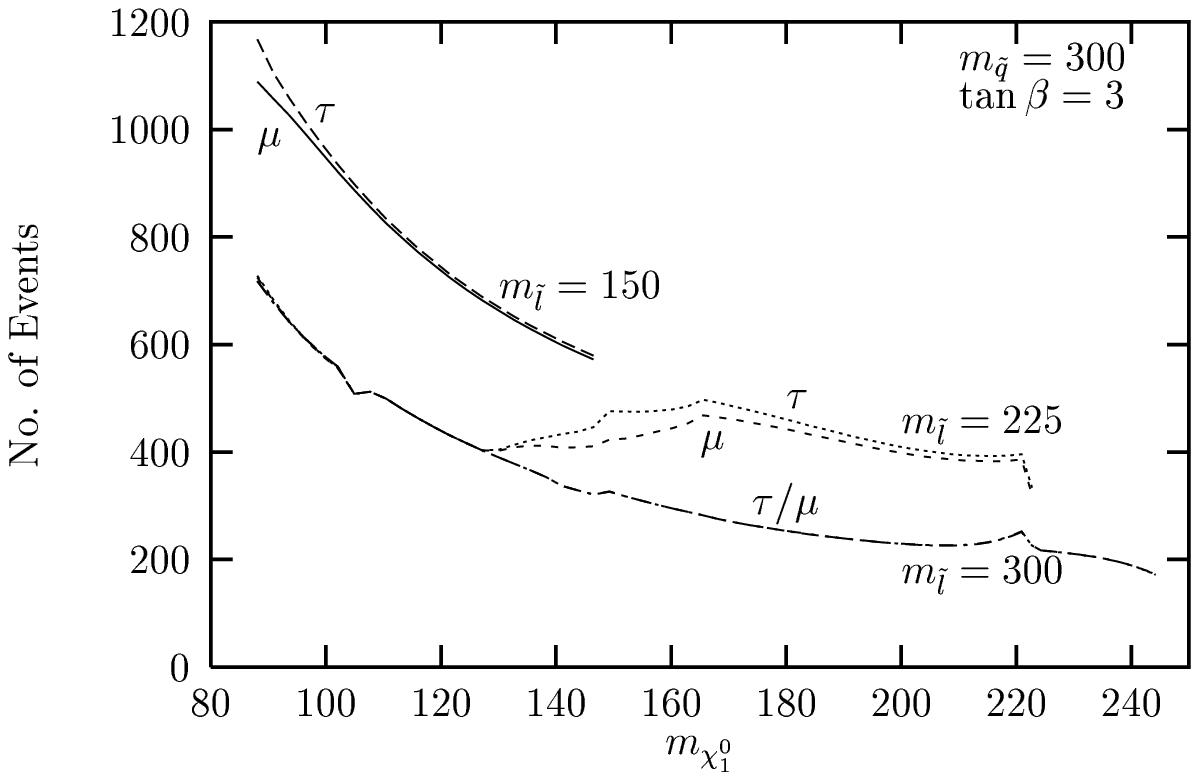,width=24cm}}
\vskip-24cm
\hskip-1cm\centerline{\epsfig{file=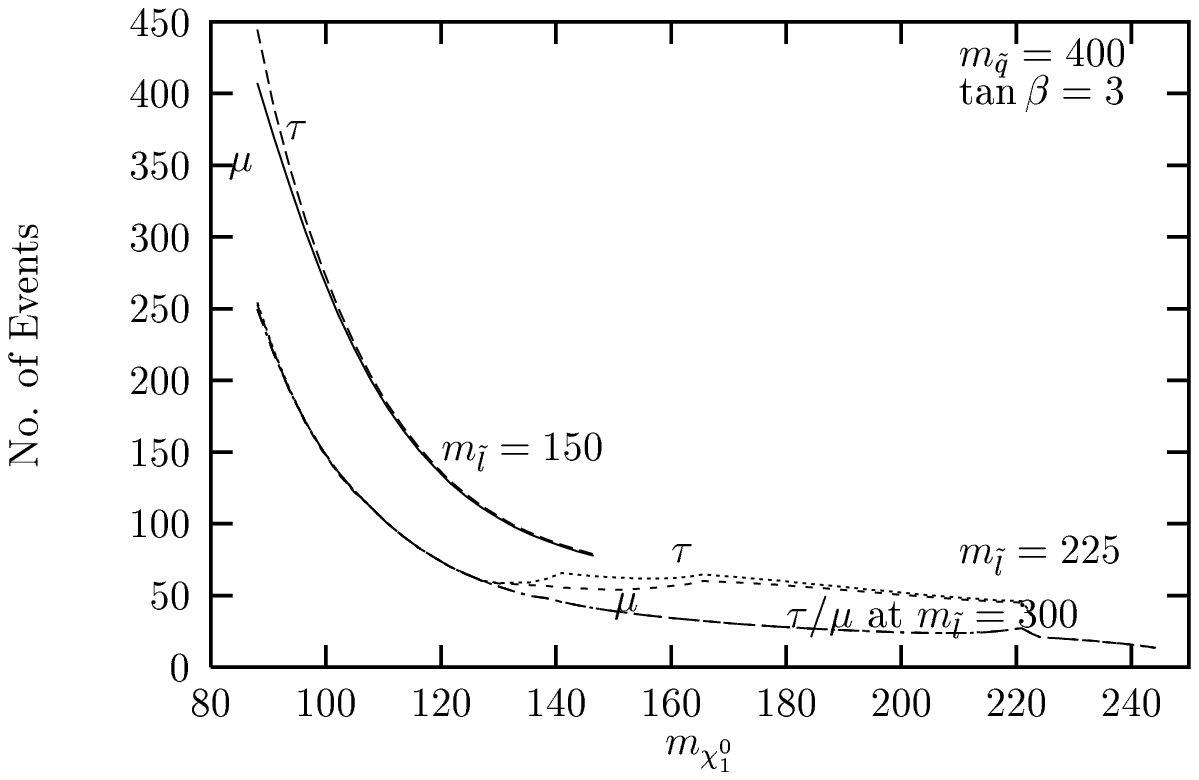,width=24cm}}
\vskip-19cm
\end{center}
{\small Figure 1: The predicted numbers of like-sign dimuon and ditau events
(indicated by the labels $\mu$ and
$\tau$ on the respective curves) with a real $W$ and a displaced vertex, 
as a function of mass of the lightest neutralino (LSP) at the upgraded Tevatron
with an integrated luminosity of $10 ~ fb^{-1}$. 
Three values of the low energy slepton mass parameter (in GeV) have been used. 
Other parameters are : $\mu'=-250$ GeV, $\tan \beta=3.$
The degenerate squark mass $m_{\tilde{q}}$ is 300 GeV
in the upper panel, 400 GeV in the lower one. 
Notice the different vertical scale.}
\end{figure}

\newpage

\begin{figure}[htb]
\begin{center}
\vskip-10cm
\hskip-1cm\centerline{\epsfig{file=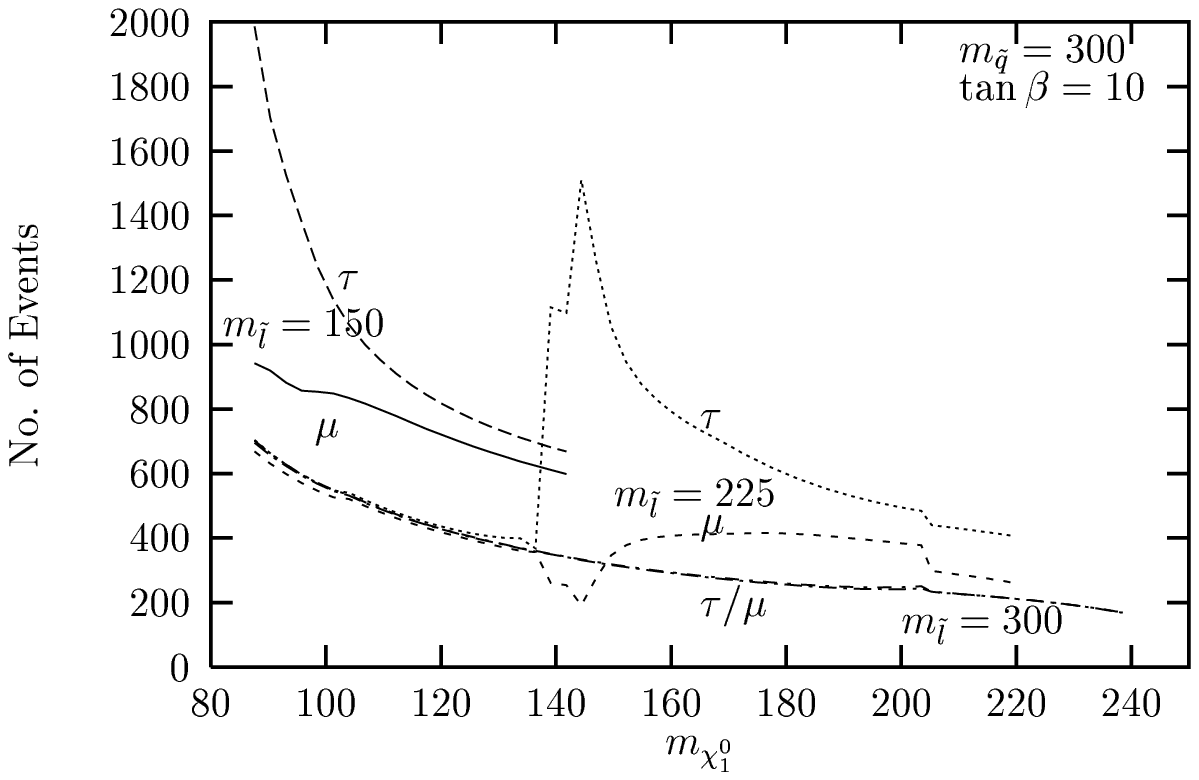,width=24cm}}
\vskip-24cm
\hskip-1cm\centerline{\epsfig{file=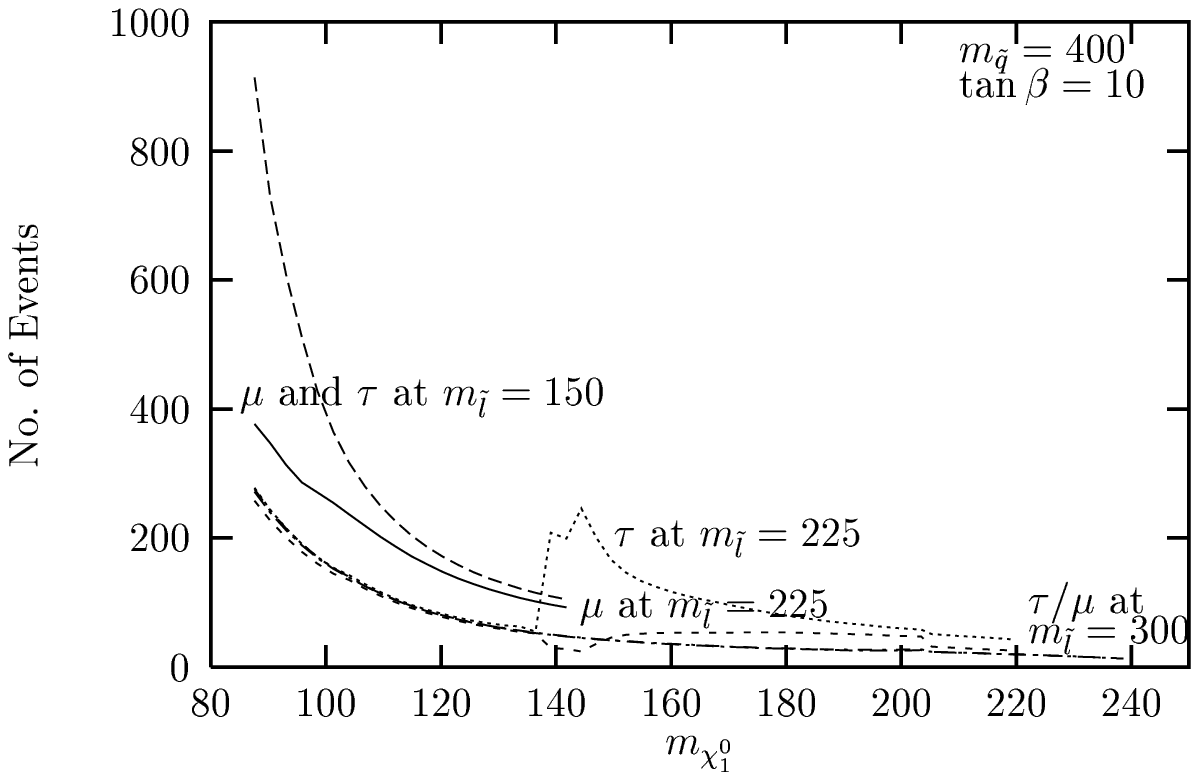,width=24cm}}
\vskip-19cm
\end{center}
\centerline{{\small Figure 2: Same as in Figure 1, 
but with $\tan \beta = 10.$}} 
\end{figure}

\newpage

\begin{figure}[htb]
\begin{center}
\vskip-10cm
\hskip-1cm\centerline{\epsfig{file=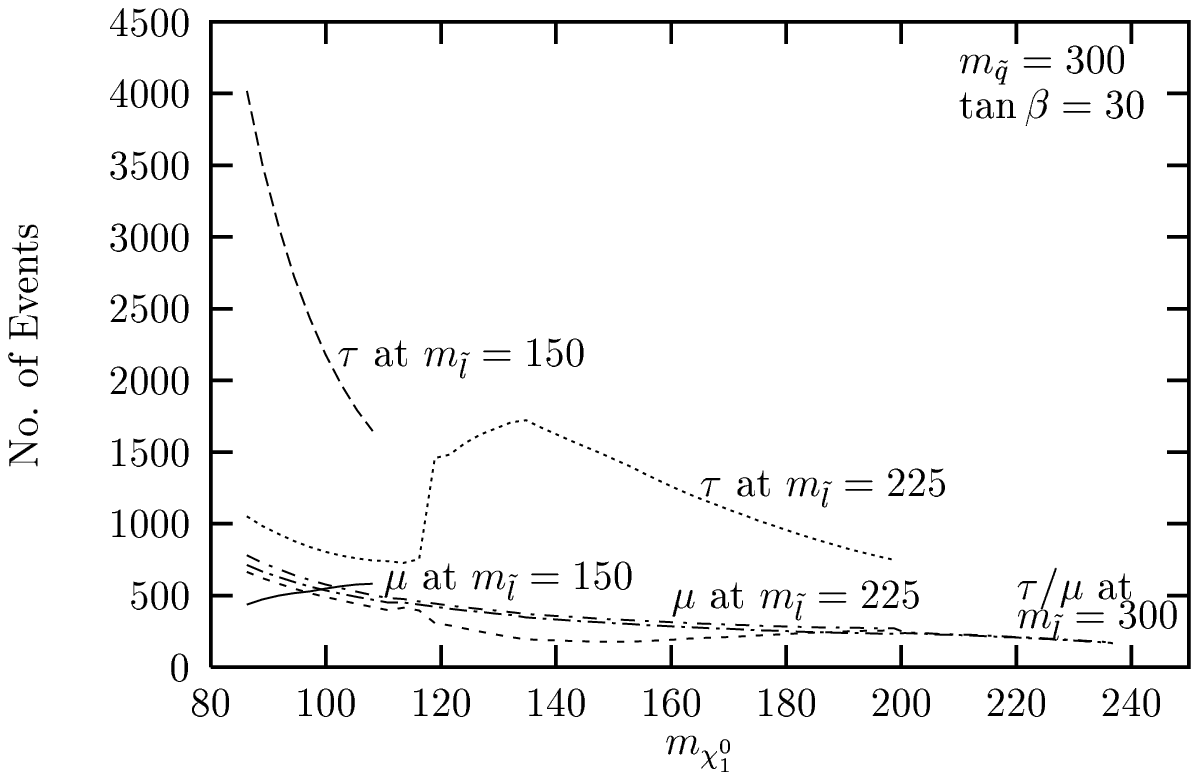,width=24cm}}
\vskip-24cm
\hskip-1cm\centerline{\epsfig{file=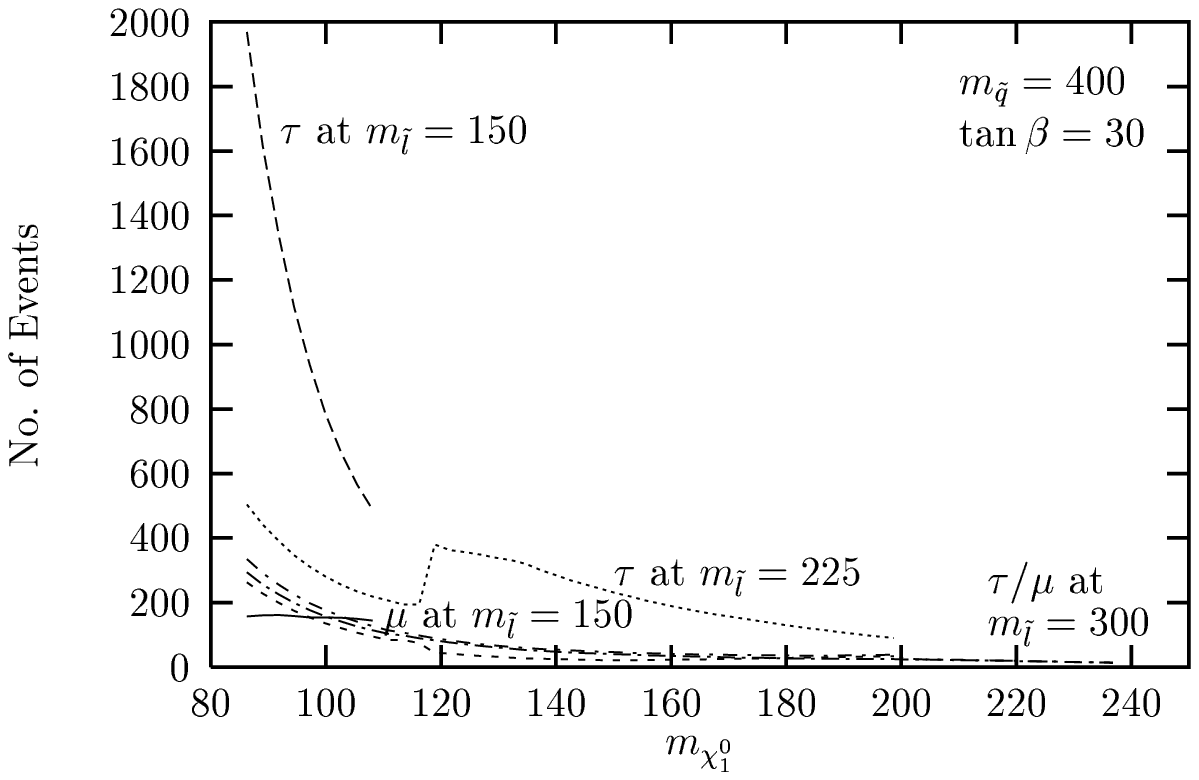,width=24cm}}
\vskip-19cm
\end{center}
\centerline{{\small Figure 3: Same as in Figure 1, 
but with $\tan \beta = 30.$} }
\end{figure}

\newpage

\begin{figure}[htb]
\begin{center}
\vskip-8cm
\hskip-1cm\centerline{\epsfig{file=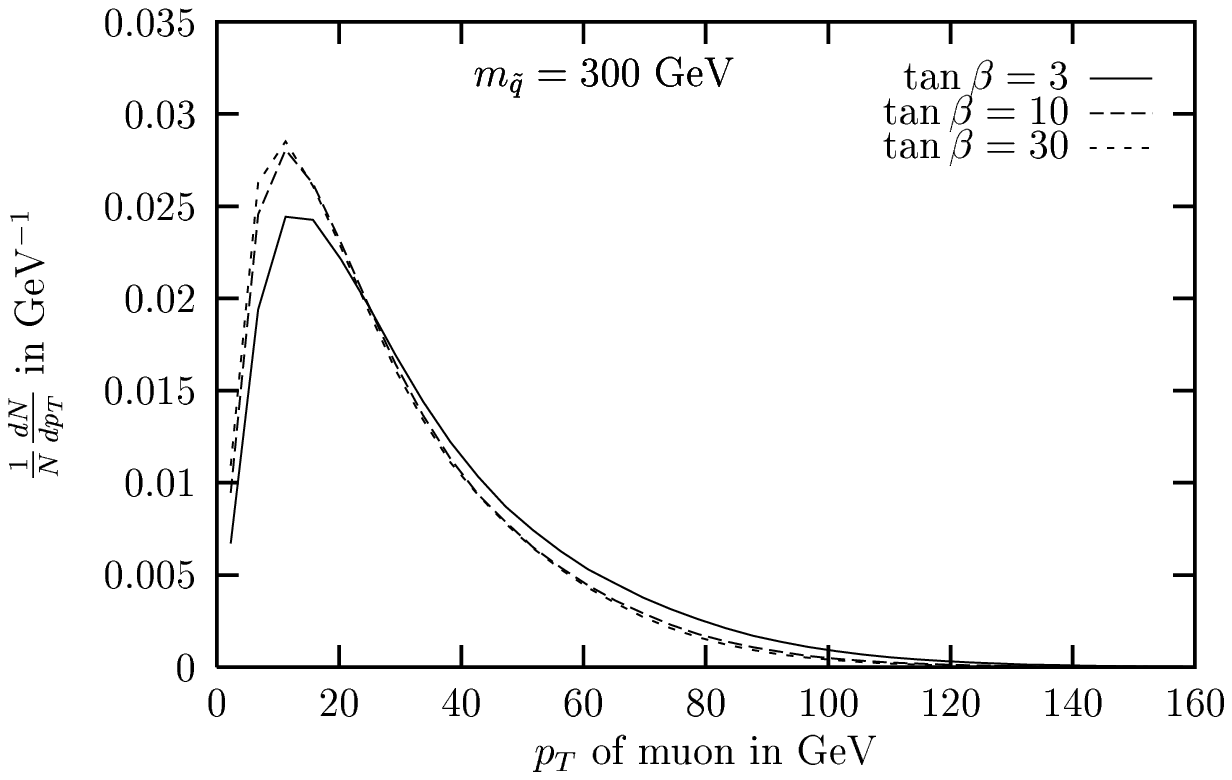,width=24cm}}
\vskip-24cm
\hskip-1cm\centerline{\epsfig{file=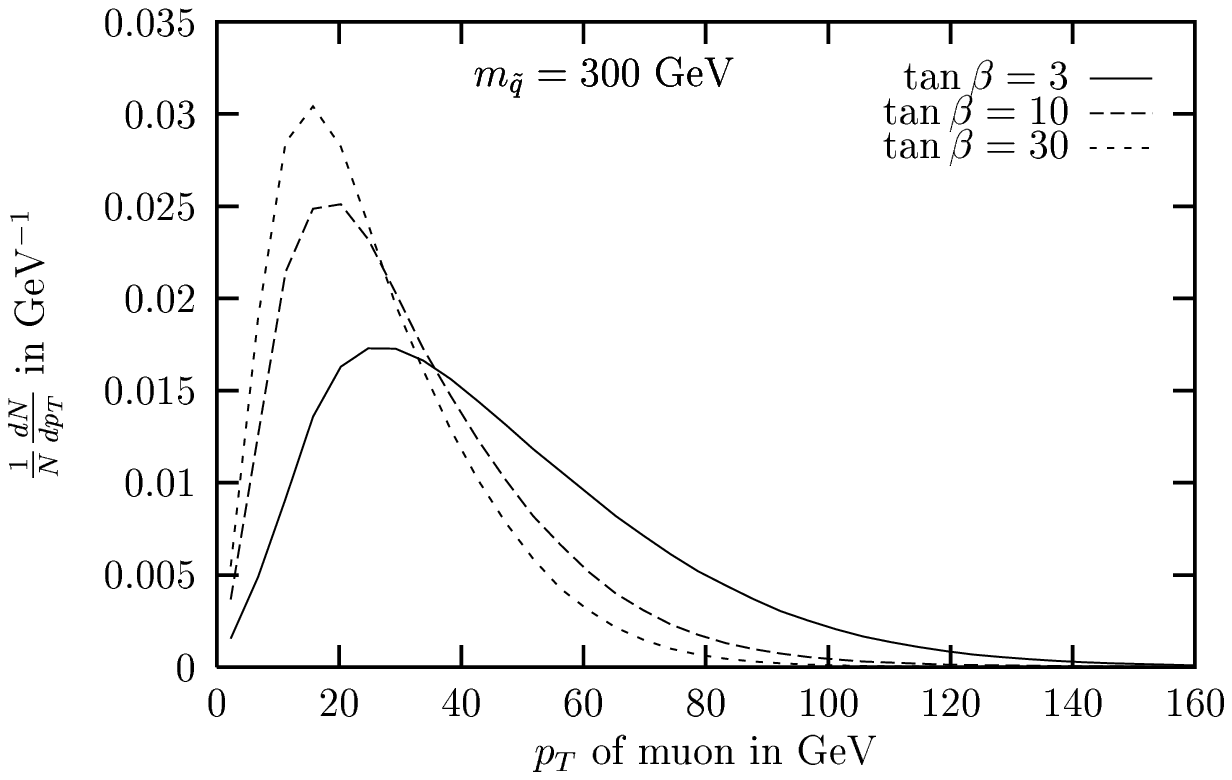,width=24cm}}
\vskip-19cm
\end{center}
{\small Figure 4:
$p_T$ distributions of muons for various values of $\tan \beta$.
The upper panel shows the distributions for muons from a displaced 
vertex due to $R$-parity violating decays of the lightest neutralino
(the LSP). The bottom panel illustrates those for muons originating from
usual MSSM cascades. 
The values of the MSSM parameters used are 
$m_{\tilde q}=300$ GeV, $m_{{\tilde l},{\tilde \nu}} \approx 150$ GeV,
$m_{\chi_1^0}=100$ GeV, $\mu'=-250$ GeV.}  
\end{figure}


\begin{thebibliography}{99}

\bibitem{neutexp} 
Super-Kamiokande Collaboration, Y. Fukuda et. al., 
Phys. Rev.Lett. 81 (1998) 1562;
MACRO Collaboration, M. Ambrosio et al., Phys. Lett. B 434 (1998) 451;
SOUDAN2 Collaboration, W.W.M. Allison et al., Phys. Lett. B 449 (1999) 137; 
CHOOZ Collaboration, M. Apollonio et al., Phys. Lett. B 420 (1998) 397; 
Phys. Lett. B 466 (1999) 415.

\bibitem{mrv} B. Mukhopadhyaya, S. Roy and F. Vissani, Phys. Lett.
B443 (1998) 191.

\bibitem{habkane}  For reviews, see, e.g., H.E. Haber and G.L. Kane, Phys.
Rep.  117 (1985) 75; G. Kane(ed.), Perspectives on
Supersymmetry (World Scientific).

\bibitem{rpv} G. Farrar and P. Fayet, Phys. Lett. B 76 (1978) 575;
C.S. Aulakh and R. N. Mohapatra, Phys. Lett. B 119 (1982) 136;
F. Zwirner, Phys. Lett. B 132 (1983) 103; L.J. Hall and
M. Suzuki, Nucl. Phys. B 231 (1984) 419; G. Ross and J.W.F. Valle,
Phys. Lett. B 151 (1985) 375; S. Dawson, Nucl. Phys. B 261 (1985) 297.

%\bibitem{barger-dreiner} V. Barger, G. Giudice and T. Han, Phys. Rev. D 40 
%(1989) 2987; H. Dreiner, hep-ph/9707435.

\bibitem{BILI} 
V. Barger, G. Giudice and T. Han, Phys. Rev. D 40 (1989) 2987;
A. Joshipura and M. Nowakowski, Phys. Rev. D 51 
(1995) 2421; {\it ibid} (1995) 5271; F. Vissani and A.Yu. Smirnov,
Nucl. Phys. B 460 (1996) 37;
R. Hempfling, Nucl. Phys. B 478 (1996) 3; 
T. Banks, Y. Grossman, E. Nardi, and Y. Nir,  Phys. Rev. D 52 (1996) 5319; 
B. de Carlos and P.L. White, Phys. Rev. D 54 (1996) 3424;
H. P. Nilles and N. Polonsky, Nucl. Phys. B 484 (1997) 33;
B. de Carlos and P.L. White, Phys. Rev. D 55 (1997) 4222;
E. Nardi, Phys. Rev. D 55 (1997) 5772;
S. Roy and B. Mukhopadhyaya, Phys. Rev. D 55 (1997) 7020;
A. Akeroyd, M.A. D\'\i az, J. Ferrandis, M.A. Garc{\'\i}a-Jare\~no, and
J. W. F. Valle, Nucl. Phys. B 529 (1998) 3;
M.A. D\'\i az, J. Ferrandis, J.C. Rom\~ao, and J.W.F. Valle, Phys. Lett. B 453
(1999) 263;
Tai-fu Feng, hep-ph/9808379; M.A. D\'\i az, E.
Torrente-Lujan, J.W.F. Valle, Nucl. Phys. B 551 (1999) 78; J. Ferrandis,
Phys. Rev. D 60 (1999) 095012;
M. Hirsch and J.W.F. Valle, Nucl. Phys. B 557 (1999) 60; C-H.
Chang and T-F. Feng, Euro. Phys. J. C 12 (2000) 137; 
F. de Campos et al., hep-ph/9903245;
R. Hempfling, hep-ph/9702412.

\bibitem{chun} E.J. Chun, S.K. Kang, C.W. Kim, and U.W. Lee,
Nucl. Phys. B 544 (1999) 89; 
V. Bednyakov, A. Faessler, and S. Kovalenko,
Phys. Lett. B 442 (1998) 203; A. S. Joshipura and S.K. Vempati,
Phys. Rev. D 60 (1999) 095009; O. Kong, Mod. Phys. Lett. A 14 (1999) 903; 
D.E. Kaplan, A.E. Nelson, JHEP 0001 (2000) 033.




\bibitem{SVEV} 
I-H. Lee, Phys. Lett. B 138 (1984) 121; 
Nucl. Phys. B 246 (1984) 120;
D.E. Brahm, L.J. Hall, S.D.H. Hsu, Phys. Rev. D42 (1990) 1860;
F. de Campos, M.A. Garc{\'\i}a-Jare\~no, A.S. Joshipura,
J. Rosiek, and J.W.F. Valle, Nucl. Phys. B 451 (1995) 3.


\bibitem{datta} A. Datta, 
B. Mukhopadhyaya and S. Roy, Phys. Rev. D 61 (2000) 055006. 

\bibitem{bahcall} J.N. Bahcall, P.I. Krastev, A.Yu. Smirnov,
Phys. Rev. D 58 (1998) 096016 
and references therein; Phys. Rev. D 60 (1999) 093001.

\bibitem{bakira} M. Drees, S. Pakvasa, X. Tata, and T. ter Veldhuis, Phys.
Rev. D 57 (1998) 5335; R. Adhikari and G. Omanovic, Phys. Rev. 
D 59 (1999) 073003;
S. Rakshit, G. Bhattarchyya, and A.  Raychaudhuri, Phys. Rev.
D 59 (1999) 091701.


\bibitem{BILEX} M. Nowakowski and A. Pilaftsis, Nucl.  Phys. B 461 (1996) 19;
A. Faessler, S. Kovalenko and F. \v{S}imkovic, Phys. Rev. D 58 (1998) 055004;
M. Bisset, O.C.W. Kong, C. Macesanu, L.H. Orr, hep-ph/9811498;
S.Y. Choi, E. J. Chun, S.K. Kang, J.S. Lee, 
Phys. Rev. D 60 (1999) 075002;
E. J. Chun, J.S. Lee, Phys. Rev. D 60 (1999) 075006;
E. J. Chun, S.K. Kang, Phys. Rev. D 61 (2000) 075012.


\bibitem{dpmona} 
See, for example,  
D.P. Roy, Phys. Lett. B 283 (1992) 270; 
H. Dreiner, M. Guchait and D.P. Roy, Phys. Rev. D 49 (1994) 3270;
H. Baer, C. Kao and X. Tata, Phys. Rev. D 51 (1995) 2180; 
M. Guchait and D.P. Roy, Phys. Rev. D 54 (1996) 3276.
Similar considerations also apply to models with spontaneous R-parity
violation. See, for example, 
M.C. Gonzalez-Garcia, J.C. Romao and J.W.F. Valle,  Nucl. Phys. B 391 (1992)
100.

\bibitem{cdf} See, for example, CDF Collaboration, F. Abe et al., 
Phys. Rev. Lett. 83 (1999) 2133.

\bibitem{d0lim} D\O ~ Collaboration, B. Abbott {\it et al.}, Phys. Rev. Lett.
83 (1999) 4937.

\bibitem{tata} 
R. Arnowitt and P. Nath, Mod. Phys. Lett. A 2 (1987) 331;
X. Tata, Lectures presented at the 1995 Theoretical 
Advanced Study Institute, University of Colorado, Boulder, USA, UH-511-833-95. 

\bibitem{mssm0} 
The Minimal Supersymmetric Standard Model: Group Summary
Report, A. Djouadi {\it et al.}, hep-ph/9901246; 
``Supersymmetry : Where it is and How to find it'' X. Tata, 
hep-ph/9510287; 
Particle Data Group, 
European Physical Journal {\bf C3} (1998) 1.

\bibitem{pdflib} H. Plothow-Besch, Int. J. Mod. Phys. A 10 (1995) 2901.

\bibitem{spira-berger} 
W. Beenakker, R. H\"{o}pker, M. Spira and P.M. Zerwas, 
Nucl. Phys. B 492 (1997) 51; 
E.L. Berger, M. Klasen and T. Tait, Phys. Rev. D 59 (1999) 074024;
M. Spira, hep-ph/9711408.

\bibitem{et12} See, for example, 
CDF Collaboration, F. Abe {\it et al.}, Phys. Rev. Lett. 83 (1999) 2133;
D\O ~ Collaboration, B. Abbott {\it et al.}, Phys. Rev. Lett. 83 (1999) 4476. 

\bibitem{L3stau} L3 Collaboration, M. Acciarri et al., Phys. Lett. B 456
(1999) 283.

\end{thebibliography}
\end{document}